%Paper: hep-th/9402135
%From: Ambar Sengupta <sengupta@marais.math.lsu.edu>
%Date: Wed, 23 Feb 94 17:48:53 -0600

%TeXfile

\magnification=\magstep 1

\def\a{\alpha}

\def\bull{\vrule height 1.5ex

width 1.4ex depth -.1ex}% square bullet

\def\ca{{\cal A}}
\def\cc{{\cal C}}
\def\cf{{\cal F}}
\def\cg{{\cal G}}

\def\e{{\epsilon}}

\def\g{\gamma}

\def\ha{H^{(1)}}
\def\hb{H^{(2)}}
\def\inv{^{-1}}
\def\la{\langle}

\def\o{\omega}
\def\O{\Omega}
\def\ov{\overline}
\def\ovo{{\ov\O}}

\def\pr{\partial}

\def\ra{\rangle}

\def\S{\Sigma}

\def\Th{\Theta}

\def\ug{{\underline g}}

\def\RefHead{\vglue 24pt \hfil {\bf REFERENCES} \hfil

\vglue 12pt}

\def\References#1#2{
{ \advance\leftskip by 1in
\noindent
\llap{\hbox to .5in{[#1] \hfil}}#2

}}

\magnification=\magstep1

\def\Title#1#2{\vglue .5in

\centerline{\bf   #1} \vskip .5pc
\centerline{\bf  #2} \vskip 1.25pc}

\def\Author#1{\vskip .5pc
\centerline{#1}}
\def\Address#1{
\centerline{\it #1}  }

\def\Abstract#1{\vglue .3in

{\parindent=.7in
\narrower
{\noindent{\bf Abstract.} \enskip #1}

}
\vglue .5in}

\Title{The Semiclassical Limit of the Two  Dimensional}
{Quantum Yang-Mills Model}
\Author{Christopher King}
\Address{Department of Mathematics}
\Address{Northeastern University}
\Address{Boston, MA 02115}
\Author{and}
\Author{Ambar Sengupta\footnote*{
Research supported in part by  LEQSF Grant RD-A-08.}}
\Address{Department of Mathematics}
\Address{Louisiana State University}
\Address{Baton Rouge, Louisiana 70803-4918}
\vglue 12pt

\Abstract{We relate the semiclassical limit of the
quantum Yang-Mills partition function on a compact oriented surface

 to the symplectic volume of
the moduli space of flat connections, by
using an explicit

expression for the symplectic form.
This gives an

independent proof of some recent

results of Witten and Forman.}

\vskip 0.5in

\noindent{\bf 1. Introduction}
\par
The quantised Yang-Mills model on a compact
two-dimensional manifold has been the subject
of much investigation recently [F1, F2, Fo, Se1-5,

W1, W2]. Witten and Forman
have studied the semiclassical limit of
the quantum partition function $Z_{YM}$.
In this limit the infinite dimensional integral
over the space of connections reduces to a finite
dimensional integral over the

quotient of the space of flat connections by the group
of gauge transformations. This latter space
has a natural symplectic structure, and Witten shows that

the semiclassical limit of $Z_{YM}$ is proportional
to the total symplectic volume.

Our main goal in this paper is to use a concrete
expression for the
symplectic structure of moduli space

to give an independent understanding of

Witten's result.
Our expression for the symplectic form

is derived in [KS], where some of its
properties are also established.

We will denote by $\cal A$

the infinite dimensional space of connections

on a principal $G$-bundle $P$ over a compact  oriented  surface
$\Sigma$,

with structure

group $G$.
This space

has a symplectic structure [AB, Go],

and this structure is preserved by
the action of the group of gauge transformations
$\cal G$. The quantum partition function is
$Z_{YM}(\epsilon) =

\int_{\cal A} exp ({-1 \over 2 \epsilon}
\int ||F_{A}||^{2} ) D A$, where $F_{A}$ is the
curvature of the connection $A$.

Using methods from quantum field theory and
algebraic arguments involving torsion
Witten determines the limit

$\lim_{\epsilon \downarrow 0} Z_{YM}(\epsilon)$.

In this paper we make use of an explicit

identification of the moduli space of flat
connections on $\Sigma$ with a

quotient of a submanifold
of $G^{2g}$ where $g$ is the genus of $\Sigma$.
This leads to a concrete expression for the natural symplectic
structure on this space in terms of
these coordinates [KS].
Using this  we investigate the results of
Witten and Forman [Fo]  in terms of our coordinate
description. The semiclassical limit
 $\lim_{\epsilon \downarrow 0}
Z_{YM}(\epsilon)$ can be computed using
lattice gauge theory. The resulting expression
is  an integral over the same submanifold
of $G^{2g}$ which we use to evaluate the symplectic
form.  Witten's identity leads us to compare
this integrand with the symplectic volume form, and hence to

a relation between certain

determinants. We verify this relation, and
hence give an independent way of understanding
Witten and Forman's result. Along the way we give
a simple proof of the non-degeneracy of the symplectic form.

The paper is organised as follows. In section 2 we

specify the identification of the

moduli space with a quotient $\cf/G$ of a submanifold of
$G^{2g}$, explain the explicit formula for the
symplectic form, and quote from [KS] a number of properties for
this form.
In section 3 we prove identities for certain
determinants on $\cf$, related to the symplectic
volume form on the moduli space. We also

prove non-degeneracy of the 2-form giving the symplectic
structure on moduli space.
In section 4
we describe the semiclassical limit of
quantum Yang-Mills theory, and prove an identity which
relates this to the symplectic volume of moduli space.
This reproduces the results of Witten and Forman
on the smooth part of moduli space.

\vskip .15in

\noindent{\bf 2. Notation, Definitions and Previous
Results}
\vskip .15in

We begin with a brief description of the setup for the

two dimensional quantum Yang-Mills model,
and a summary of relevant results from [KS].
Let $\S$ be  a

compact oriented surface (without boundary)  of genus $g\geq 1$.

Let $G$ be a compact, connected,  simply connected
Lie group. The Lie algebra of $G$
will be denoted $\ug$, and equipped with a fixed $Ad-$invariant

inner-product $\la\cdot,\cdot\ra_{\ug}$.

Let $\pi :P\to\S$ be a principal

$G-$bundle over $\S$, and let $\ca$
denote the set of all connections on $P$.

Let $\cal G$ be the set of all smooth bundle

automorphisms $\phi :P\to P$ (i.e. $\phi$ is a

diffeomorphism commuting with the $G-$action) such

that $\pi\phi=\pi$.  Then $\cal G$ is a group under composition
and acts on $\ca$ by pullbacks : $(\phi,\o)\mapsto \phi^*\o$.

A  connection

$\o$ is  said to be {\it flat}  if

$d\o(X,Y) + [\o(X),\o(Y)]=0$, for

every $X,Y\in T_pP$ and every $p\in P$. The space of all
{\it flat connections} on $P$ will be denoted $\ca^{fl}$.

The action  of $\cg$
on $\ca$ maps $\ca^{fl}$ onto itself and we denote the

quotient  by $\cc^{fl}$ :
$$\cc^{fl}=\ca^{fl}/\cg.$$

A tangent vector to $\ca$

is the difference of two connections and is thus a $\ug-$valued

$1-$form $\Th$ on $P$ which commutes with the (right) action
of $G$ and which annihilates the vertical subspace.
The set of all such $1-$forms may be taken to

be the vector space of tangent vectors to $\ca$.

If $\Th_1$ and

$\Th_2$ are two such tangent vectors then it is easily verifiable

that $\la\Th_1\wedge\Th_2\ra$ (which is the exterior product combined
with the inner-product pairing in $\ug$) is $\pi^*$ of  a (unique)

$2-$form on

$\S$; integrating this $2-$form over $\S$ we obtain a

number which we

denote $\O(\Th_1,\Th_2)$.

Then  $\O$, thus defined, is a symplectic

form on $\ca$.  The restriction of $\O$ to $\ca^{fl}$ induces

a 2-form on the quotient  $\cc^{fl}$.

A flat connection is described by its holonomy around a
set of loops generating the fundamental group of $\Sigma$.
This leads to the following coordinate description of
$\cc^{fl}$ (see [KS] for the derivation). Define
$$\cf =\{(a_1,b_1,...,a_{g},b_{g})\in G^{2g} :
{b}_{g}^{-1}{a}_{g}^{-1}{b}_g {a}_g...{b}_1^{-1}
{a}_1^{-1}{b}_1{a}_1=e\}$$

The group $G$ acts on $\cf$ by :
$$\Bigl(x,(a_1,b_1,....,a_g,b_g)\Bigr)\mapsto
(xa_1x^{-1},...,xb_gx^{-1}\Bigr).$$

There is a bijection from $\cc^{fl}$ to
$\cf/G$, defined by holonomies around the generating loops.
This is our explicit coordinate description of $\cc^{fl}$.
Before writing down the symplectic form on $\cc^{fl}$,
we observe that $\cf$ can be identified with a subset of
$G^{4g}$, as follows. Define

$$J \buildrel\rm def\over =\{1,2,5,6,....,4g-3,4g-2\}$$

Then $\cf \subset G^{2g}$

can be identified with the subset of $G^{4g}$

consisting of those elements $(\a_1,...,\a_{4g})\in G^{4g}$  for

which $\a_{j+2}=\a_j^{-1}$ for every
$j\in J$, and ${\a}_{4g}....{\a}_1 =e$.

{\it  Our notation will often

assume this identification.}  Specifically,  for

$(a_1,b_1,...,a_g,b_g)\in G^{2g}$ we set  $\a_1=a_1$, $\a_2=b_1$,

$\a_3=a_1^{-1}$, $\a_4 = b_1^{-1}$,..., $\a_{4g}=b_g^{-1}$.

Our expression for the symplectic form is derived from a
2-form $\O$ on $\cf$. This 2-form is given most conveniently
by using the identification with $G^{4g}$.
A tangent vector at $(\a_1,...,\a_{4g})\in G^{4g}$
is obtained by left translation from

$4g$ elements of $\ug$, which we write
$H_{1},...,H_{4g}$. We also define
$$s_0=e\qquad\hbox{and}\qquad  s_i = \a_i....\a_1\qquad\hbox{for

every}\qquad i\in\{1,....,4g\}$$

If $(\a_1,...,\a_{4g})$ corresponds to an element of $\cf$,
then the $H_i$ have the following relationships

between themselves :

$$H_{j+2}  = - Ad(\a_j) H_j     \qquad \hbox{ for}\qquad j\in J$$

$$Ad(s_{j+1}^{-1})H_{j+2}   = - Ad(s_{j+2}^{-1})H_j  \qquad \hbox{

for}\qquad j\in J$$

$$\sum_{i=1}^{4g} Ad(s_{i-1}^{-1})H_i  =0$$

With these definitions, it is shown in [KS] that the 2-form
$\O$ on $G^{2g}$ is given by the following expression:
$$\O(\a H^{(1)},\a H^{(2)}) =

{1\over 2}\sum_{1\leq i, k\leq 4g}

\e_{ik} \la Ad(s^{-1}_{i-1})H^{(1)}_i,

Ad(s^{-1}_{k-1})H^{(2)}_k \ra  $$
where  $\e_{ik} =1$ if $i<k$, $\e_{ik}=-1$ if $i>k$,

and $\e_{ik}=0$ if $i=k$.

It is proved in [KS] (Proposition 4.5)

that $\O$ is closed. Let $\cf^{0}$ denote the set of points
of $\cf$ at which the map

$K:  G^{2g}\to G : (a_1,b_1,...,a_g,b_g)\mapsto

c_g...c_1$, wherein $c_i=b_i^{-1}a_i^{-1}b_ia_i$,

has derivative of

full rank (=$\dim(G)$). Thus $\cf^0$, if it is non-empty,

is a smooth submanifold of $G^{2g}$ of dimension  $(2g-1)\dim(G)$.
Proposition 3.2 of [KS] proves that if $g \geq 2$,
there is a non-empty open subset $\cf^{1} \subset \cf^{0}$
such that :

(i) the isotropy group at each point of $\cf^1$ is $Z(G)$

(the center of $G$); (ii) $\cf^1/G$ has a smooth
 manifold structure and $p :\cf^1\to \cf^1/G$  is

then a smooth bundle with fiber $G/Z(G)$.

{\it  Assumption/Notation} :   We shall assume henceforth

that there exists an open dense subset
$\cf^{1} \subset \cf^{0}$ such that the isotropy
group at each $x \in \cf^{1}$ is $Z(G)$, that is
$$
\{ g \in G : gxg\inv =x \} = Z(G).$$
(In this connection see section 3 of [KS].

According to Theorem 4.3.1 in
Bredon [Br] an open dense subset of minimal
isotropy type - within each connected component - always exists,

and, by  Proposition 4.3 of [KS], this minimal isotropy group

is discrete; we are taking

it to be $Z(G)$.)

Corollary 6.2.5 of [Br] implies that
${\cf^{1}/G}$ is a $C^{\infty}$ manifold, that the
quotient map $p: \cf^{1} \rightarrow

{\cf^{1}/G}$ is smooth, and that
$\cf^{1} {\buildrel p \over \rightarrow}

{\cf^{1}/G}$ is a fiber bundle with fiber
$G/Z(G)$. Therefore each $y \in {\cf^{1}/G}$
has an open neighborhood $U$,  there is a

smooth map $s: U \rightarrow \cf^{1}$ with

$p\circ s = id$, and the map
$$
\eqalign{
\psi : \, & G/Z(G) \times U \rightarrow p\inv (U) \cr
&(g. Z(G), u) \mapsto g s(u) g\inv \cr}
$$
is a G-equivariant diffeomorphism.

It is proven in Lemma 4.1 in [KS] that the

2-form $\O$ annihilates the
subspace of $T_\a\cf$ which is tangent to the orbit of the
$G$-action.  Since $\O$ is also  invariant under the  $G-$action,

it follows that
there is a closed $2-$form  $\ovo$ on $\cf^{1}/G$ such that

$\O=p^*\ovo$, where   $p:\cf^{1}\to \cf^{1}/G$ is

the quotient map.

There are some further properties of $\O$ which we will
need in the next section. These are proved in [KS], and we
quote the results below.

\medskip

{\it 2.1. Proposition}.  Let $K : G^{2g}\to G

:(a_1,b_1,...,a_g,b_g)\mapsto c_g...c_1$, where

$c_i=b_i^{-1}a_i^{-1}b_ia_i$.

Consider any $\a=(a_1,b_1,...,a_g,b_g)

\in G^{2g}$, and let $\g_\a :G\to G^{2g} $ be the `orbit map'

$x\mapsto (xa_1x^{-1},xb_1x^{-1},...,xb_gx^{-1})$, and

$\g'_\a:\ug\to T_eG^{2g} : X\mapsto \a^{-1}.d\g_\a'(X)$.

 Let $T_\a$

denote the derivative $dK_\a$ after all vectors

are `translated to

the identity'; i.e. $ T_\a:T_eG^{2g}\to \ug:X\mapsto

K(\a)^{-1}dK_\a. (\a X).$

By a convenient abuse of notation, we will write the

adjoint $T^*_\a$

of $T_\a$ as $dK_\a^*$; thus  $dK_\a^* : \ug\to T_eG^{2g}$.

Then :
$$ker\,dK_\a^* = ker \,\g_\a'.$$

Also  writing $dK_\a^*X =

\bigl((dK_\a^*X)_1,(dK_\a^*X)_2,(dK_\a^*X)_5,...,
(dK_\a^*X)_{4g-2}\bigr)$, we have  :
$$(dK_\a^*X)_j = (f_{j-1} - f_{j+2})X $$

\medskip

{\it 2.2. Proposition} :  Let us write :
$$\O(\a \ha,\a \hb)  =\la \ha, \O_b\hb\ra_{\ug^{2g}} = -\la

\O_b\ha,\hb\ra_{\ug^{2g}},$$
wherein  $\O_ b H$ has components :
$$(\O_bH)_j = -{1\over 2} \sum_{i\in J} \bigl[

f_{j-1}(\e_{ij}f_{i-1}^{-1}-\e_{i+2,j}f_{i+2}^{-1})

 - f_{j+2}(\e_{i,j+2}f_{i-1}^{-1} - \e_{ij}f_{i+2}^{-1})\bigr]H_i$$
for $j\in J$, where  $f_i =Ad(\a_i...\a_1)$, and $f_i$ is

the identity $1$ if $i\leq 0$ or if $i>4g$.

Let  $\a=(a_1,b_1,...,a_g,b_g)\in K^{-1}(e)\subset G^{2g}$,

and  let  $K : G^{2g}\to G $, $\g_\a :G\to G^{2g}$, and   $dK_\a^* :

\ug\to T_eG^{2g}$ be as in  Proposition  2.1. Then :
$$\O_b\g'_\a = dK_\a^*.$$

\medskip

\noindent{\bf 3. The Symplectic Volume Form on Moduli Space}
\vskip .15in

To begin, we state some conventions which will be
used in this section.
The fixed Ad-invariant inner product on $\ug$ induces
a Riemannian structure on $G$, and hence on
$G^{n}$. Our convention is that we always use the
corresponding  measure  on $G^{n}$. For any
Riemannian manifold, we denote by $d vol$
the Riemannian volume measure.

If $A: V \rightarrow W$ is a linear map of inner
product spaces, then $det A$ will denote
$\pm (det(A^{\ast} A))^{1/2}$. This is the

determinant of the matrix of $A$ relative to

orthonormal bases in $V$ and in $AV \subset W$
(enlarged to yield a square matrix for $A$),
assuming that ${\rm dim} W \geq {\rm dim} V$.

If $\Lambda$ is a 2-form on an even-dimensional
inner product space $V$, we denote by $Pf(\Lambda)$
the Pfaffian of the skew symmetric matrix
$\Lambda(E_{i}, E_{j})$ where $\{E_{i} \}$ is an
orthonormal basis of $V$ (see section 12.3 of
[CFKS] for the definition and properties of the
Pfaffian). This is undetermined up to sign, but this
ambiguity will not matter in our calculations.

If $\omega$ is a 2-form on a Riemannian manifold
of dimension 2r,
$$
{\omega^{r} \over r!} = \pm Pf(\omega) \,d vol$$
where
$Pf(\omega)(x) = Pf(\omega(x))$. We adopt the
convention that $|\eta|$ is the positive measure
associated with the top form $\eta$. Therefore
$$
{|\omega^{r}| \over r!} =

|Pf(\omega)|\, d vol \eqno(3.1)$$

Henceforth we equip ${\cf^{1}/G}$ with the
natural Riemannian metric coming from the
quotient map and the metric on $\cf^{1}$, that is
if $X,Y \in T_{p(y)} ({\cf^{1}/G})$ then
$\langle X, Y\rangle =

\langle{\tilde X}, {\tilde Y}\rangle$ where
${\tilde X}$ is the unique element of
$T_{y}(\cf^{1})$ orthogonal to

${\gamma}_{y}'(\ug)$ and for which
$p_{\ast} {\tilde X} = X$, and ${\tilde Y}$ is the
corresponding element for $Y$
(recall that $\g_\a :G\to G^{2g} $ is the `orbit map'

$x\mapsto (xa_1x^{-1},xb_1x^{-1},...,xb_gx^{-1})$).

{\it 3.1. Proposition.}
For any $u \in {\cf^{1}/G}$, and any $g \in G$,
$$
det \,{\psi'}_{(g. Z(G), u)} = det \,{\gamma'}_{s(u)}$$
(recall our convention that det is defined up to
sign).

{\it Proof:}
Let $e_{1}, \dots, e_{r}$ be an orthonormal basis

of $\ug$, and $x_{1}, \dots, x_{k}$ an orthonormal
basis of

${\gamma'}_{s(u)}(\ug)^{\perp} \cap

T_{s(u)} {\cf}^{1}$. Write $x_{i} = x_{i}' + y_{i}$
where $y_{i} \in {\gamma'}_{s(u)}(\ug)$ and
$x_{i}' \in T_{s(u)}(s U)$. So
$\{p_{\ast} x_{i}' \} = \{p_{\ast} x_{i} \}$
is an orthonormal basis of $T_{u} (\cf^{1}/G)$.

Furthermore

${\psi'}_{(e. Z(G), u)}(0, p_{\ast}x_{i}) =
s_{\ast} p_{\ast}(x_{i}) =
s_{\ast} p_{\ast}(x_{i}') = x_{i}' $.
So the matrix of

${\psi'}_{(e. Z(G), u)}$ with respect to these bases is
$$
\left( \matrix{{\gamma'}_{s(u)} & \ast \cr
0 & 1 \cr} \right)$$

Thus $det \,{\psi'}_{(e. Z(G), u)} = det \,{\gamma'}_{s(u)}$.
Furthermore $\psi$ is G-equivariant and the G-actions

on the domain and range of $\psi$ are isometries, so
$det \,{\psi'}_{(g. Z(G), u)}$ is independent of $g$.
$\bull$

\medskip
Being a submanifold of $G^{2g}$, $\cf^{1}$ inherits a
Riemannian metric and hence a corresponding

volume measure. Proposition 3.1 implies the

following result.

{\it  3.2. Corollary.}
Let $f$ be a continuous function with compact support
on $\cf^{1}/G$. Then
$$
\int_{\cf^{1}} f\circ p \,d vol =
{vol(G) \over |Z(G)|} \int_{\cf^{1}/G}
f \,|det {\gamma}'|\, d vol $$
where $vol(G)$ is the total Riemannian volume of G,
and $|Z(G)|$ is the order of $Z(G)$.

\medskip
{\it  3.3. Proposition. }
$$
Pf(\overline{\Omega}) =  \,
{det {\gamma}' \over det (d K^{\ast})}
\quad {\rm on}\,\, \cf^{1} $$
(recall that both Pf and det are defined up to sign).

{\it Proof:}
Let us consider $\Omega$ as a 2-form on $G^{2g}$,
and, for $y\in\cf^1$,  write (with orthogonal summands)
$$
T_{y} G^{2g} = d K^{\ast}_{y}(\ug) \oplus
{\gamma}_{y}'(\ug) \oplus

({\gamma}_{y}'(\ug)^{\perp} \cap
T_{y}(\cf^{1}))$$

Identifying

${\gamma}_{y}'(\ug)^{\perp} \cap
T_{y}(\cf^{1})$ with $T_{p(y)}(\cf^{1}/G)$ via the
isometry $p_{\ast}$ we can write a matrix of
$\Omega$ (to be precise,

of the operator $\O_b :{\ug}^{2g}\to {\ug}^{2g}$

corresponding to $\O$, as in Proposition 2.2)

relative to the above decomposition:
$$
\Omega =

\left(\matrix{
\ast & Q & \ast \cr
-Q^{\ast} & 0 & 0 \cr
\ast & 0 & p^{\ast} \overline{\Omega}\cr} \right)$$
where $Q: {\gamma}'(\ug) \rightarrow

d K^{\ast}(\ug)$, $X \mapsto {\Omega}_{b} X$, and
${\Omega}_{b}$ is defined in Proposition 2.2.
Also

$p^{\ast} \overline{\Omega}$ is restricted to
${\gamma}_{y}'(\ug)^{\perp} \cap
T_{y}(\cf^{1})$. Then Proposition 2.2
implies that

$$
det Q \, det (\gamma') = det (d K^{\ast})$$

Furthermore by calculating determinants

and noting that $Pf(p^{\ast} \overline{\Omega})
= Pf(\overline{\Omega})$,
we have
$$
Pf(\Omega) = det Q \,\, Pf(\overline{\Omega})$$
and hence
$$
Pf(\overline{\Omega}) = Pf(\Omega) \,
{det \gamma' \over det (d K^{\ast})}.$$

The proof is now completed by recalling that
$Pf(\Omega)^{2} = det (\Omega)$, and
using  Lemma 3.4 below.

$\bull$

{\it  3.4. Lemma.}  $det(\Omega)=1$.

{\it Proof} :  Let $\O''= \O_b -{1\over 2}(dK)^*(dK)$.

Here $\O_b :{\ug}^{2g}\to {\ug}^{2g}$ is the operator

corresponding to $\O$, as in Proposition 2.2,

and  we have written

$dK$ to mean $K^{-1}dK :{\ug}^{2g}\to \ug$,

at any point on $G^{2g}$ (and $(dK)^*$ is its adjoint).

As in Proposition 3.3 we will not   make a distinction

between $\O$,  $\O_b$ and the corresponding matrix, and

similarly for $\O''$.

Let  $(\O''H)_j=\sum_{k\in J}C_{jk}H_k$,

where $j$ runs over $J=\{1,2,5,6,...,4g-3,4g-2\}$,

and $H=(H_j)_{j\in J}\in {\ug}^{2g}$.

By   Proposition 2.2 (which gives a formula for the matrix

for  $\O_b$) and Proposition 2.1 (describing $(dK)^*$),

we see that $C_{jk}=0$ for $j,k\in J$ with $j+2<k$.

Viewing the matrix for $\O''$ as a $(2g)\times (2g)$

matrix  whose entries are $ (2\dim G)\times (2\dim G)$ blocks,

we thus see that each block  above the main diagonal

(consisting of blocks) is zero.

Therefore, $det(\O'')=\prod_{j\in J'}det(D_j)$,

wherein $D_j$ is the diagonal block starting with $C_{j j}$

as its first  (`top left') entry, and $J'=\{1,5,9,...,4g-3\}$.

Using Proposition 2.2 and the expressions for $dK$ and $(dK)^*$,

we  have :

$$D_j\buildrel \rm def \over =\, \left(\matrix{C_{jj} & C_{j\,j+1}\cr
C_{j+1\,j } & C_{j+1\,j}\cr}\right) =\,\left(\matrix{
f_{j+2}f_{j-1}^{-1} -1& f_{j+2}f_j^{-1}\cr
f_{j+3}(f_{j-1}^{-1} - f_{j+2}^{-1}) - f_jf_{j-1}^{-1} &
f_{j+3}f_j^{-1} -1\cr}\right).$$

Recall that $f_j=Ad(\a_j...\a_1)$, with usual notation.

For notational convenience we will drop $Ad$;

thus $f_j=\a_j...\a_1$. Recall also that $\a_{j+2}=\a_j^{-1}$.

To simplfy the notation even further, we will write $a=\a_j $,

and $b=\a_{j+1}$, for any fixed $j\in J'$. Then :

 $$D_j  =\, \left(\matrix{a^{-1}ba -1  & a^{-1}b\cr
b^{-1}a^{-1}ba -b^{-1}-a & b^{-1}a^{-1}b -1\cr}\right).$$

This matrix can be factorised as follows:

$$D_{j} =\, \left(\matrix{a^{-1}  & 0\cr
0 & b^{-1}\cr}\right)
\, \left(\matrix{0  & 1\cr
-1 & a^{-1} -1\cr}\right)
\, \left(\matrix{1  & 0\cr
b -1 & 1\cr}\right)
\, \left(\matrix{a  & 0\cr
0 & b\cr}\right)\,,$$
and this  implies that
$det(D_j)=1$.
Therefore :
$$det (\O''  )=1.$$

Let $y\in\cf^1$, and recall from the proof of Proposition 3.3

the decomposition

$$ T_{y} G^{2g} = d K^{\ast}_{y}(\ug) \oplus
{\gamma}_{y}'(\ug) \oplus

({\gamma}_{y}'(\ug)^{\perp} \cap
T_{y}(\cf^{1})).$$

Since $(dK)^*(dK)$ vanishes on $T_{y}\cf^{1}$  and has range
contained in  $(dK)^*(\ug)$, the matrix of
${1\over 2}(dK)^*(dK)$, relative to orthonormal bases in the

decomposition of $T_yG^{2g}$ given above, has a non-zero entry
only in the top left corner (corresponding to the subspace
$d K^{\ast}_{y}(\ug)$).

Examining the two matrices for $\O$ and $\O''$ we now see

that their determinants are equal (the common value being

$(det Q)^2  det(p^*\ovo)$).  Since we already know that

$det(\O'')=1$,  we  conclude that $det(\O)=1$. $\bull$

Proposition 3.3 implies that

$\overline{\Omega}$ is
non-degenerate, and hence symplectic, and that
therefore $\cf^{1}/G$ is orientable. In [Go],

Goldman views the symplectic structure by means

of a cup product and duality in group cohomology, and

points out that the non-degeneracy of the symplectic structure,

in that setting, is a consequence of  the   non-degeneracy

of the cup-product pairing known in group cohomology.

Since $K$ is G-equivariant with the natural conjugation
actions of G on its domain and range, and these
actions are isometries, it follows that
$det (d K^{\ast})$ is a G-invariant function on

$\cf^{1}$. A similar argument shows that

$det (\gamma')$ is G-invariant  on $\cf^{1}$.

{\it  3.5. Proposition.}
Let f be a continuous function with compact support on
$\cf^{1}/G$. Let $r = (g-1) \dim G$. Then
$$
\int_{\cf^{1}} f \circ p\,
{1 \over |det (d K^{\ast})|} \,d vol =
{vol(G) \over |Z(G)|}

\int_{\cf^{1}/G} f \,{| \,{\overline{\Omega}}^{r}| \over
r!}
$$

{\it Proof:}
 From Proposition 3.3 and Corollary 3.2,
$$
\eqalign{
\int_{\cf^{1}} f \circ p \,
{1 \over |det (d K^{\ast})|} \,d vol & =
\int_{\cf^{1}} f \circ p\,
{|Pf(\overline{\Omega})| \over |det (\gamma')|} \,d vol \cr
& =
{vol(G) \over |Z(G)|}

\int_{\cf^{1}/G} f \,|Pf({\overline{\Omega}})| \,d vol \cr}
$$

The stated result follows from this and  equation (3.1).
$\bull$

\medskip

\noindent{\bf 4. The Semiclassical Limit of the
Yang-Mills model}
\vskip .15in

We turn now to  an examination  and explanation of

some of  the results in [W1] and [Fo] in terms of our results.

The quantum partition function on $\Sigma$ is
$$
Z_{YM}(g, t) =
\int_{G^{2g}} Q_{t}(K(x)) \,d vol(x)$$
where g is the genus of $\Sigma$.

The heat kernel $Q_{t}(x)$
is defined as the fundamental

solution of the heat equation
${\pr Q_{t} \over \pr t} = {1 \over 2} \triangle Q_{t}$,
where $\triangle$ is the Laplacian on $G$, and
$\int_{G} Q_{t} d vol =1 $.

In [W1] Witten
shows that
$$
\lim_{t \downarrow 0} Z_{YM}(g, t) =

{c \over |Z(G)|} \int_{\cf /G}

{{\overline{\Omega}}^{r} \over
r!}\eqno(4.1)$$
where $c$ is a number depending on the inner
product in $\ug$, but not depending on the genus
of $\Sigma$. Different conventions about the
definition of $Q_{t}$ and the measure on G lead to
different values of $c$. Scaling the metric shows
that $c$ is proportional to $vol(G)$.

Let f be a continuous function on $G^{2g}$ which is
zero in  a neighborhood of $\cf - \cf^{0}$. Then
a change of variables and the properties of the heat
kernel imply that
$$
\lim_{t \downarrow 0} \int_{G^{2g}} f\,
Q_{t}(K(x)) \,d vol(x) =

\int_{\cf^{0}} f \,{1 \over |det (d K^{\ast})|} \,d vol
$$

Suppose in addition that f is a continuous

G-invariant function on

$G^{2g}$ which vanishes in a neighborhood of
$\cf - \cf^{1}$. Let ${\tilde f}$ be the function

on $\cf^{1} /G$ induced by f, so that

$f={\tilde f} \circ p $. Then Proposition 3.5 implies that
$$
\lim_{t \downarrow 0} \int_{G^{2g}} f(x)\,
Q_{t}\bigl(K(x)\bigr) \,d vol(x) =

{vol(G) \over |Z(G)|} \int_{\cf^{1} /G}

{\tilde f} \,{|{\overline{\Omega}}^{r}| \over
r!} \eqno(4.2)$$

This is a ``local" version of Witten's result (4.1), which
avoids the difficulties associated with the singular
subset of $\cf$. Assuming that

$|c|  = vol(G)$ it agrees (up to sign) with the

corresponding results of Forman [Fo].

\vskip 0.5in

\RefHead

\References{AB}{M. Atiyah and R. Bott, {\it
The Yang-Mills Equations over Riemann Surfaces},
Phil. Trans. R. Soc. Lond. {\bf A 308},

523-615 (1982)}
\References{Br}{G. Bredon, {\it Introduction to
Compact Transformation Groups},

Academic Press (1972)}
\References{CFKS}{H. L. Cycon, R. G. Froese,

W. Kirsch and B. Simon, {\it Schr\"odinger Operators},
Springer-Verlag (1987)}
\References{F1}{D. Fine, {\it Quantum Yang-Mills
on the Two-Sphere}, Commun. Math. Phys. {\bf 134},
273-292 (1990)}
\References{F2}{D. Fine, {\it Quantum Yang-Mills
on a Riemann Surface}, Commun. Math. Phys. {\bf 140},
321-338 (1991)}
\References{Fo}{R. Forman, {\it Small volume
limits of 2-d Yang-Mills}, Commun. Math. Phys. {\bf 151},
39-52 (1993)}
\References{Go}{W. Goldman, {\it The Symplectic
Nature of Fundamental Groups of Surfaces},
Adv. Math. {\bf 54}, 200-225 (1984)}
\References{KS}{C. King and A. Sengupta, {\it An Explicit
Description of the Symplectic Structure of Moduli Spaces
of Flat Connections}, preprint 1994}
\References{Se1}{A. Sengupta, {\it The Yang-Mills
measure for ${\rm S}^{2}$}, J. Funct. Anal. {\bf 108},
231-273 (1992)}
\References{Se2}{A. Sengupta, {\it The Semiclassical
Limit of  Gauge Theory  on ${\rm S}^{2}$}, Commun. Math, Phys. {\bf
147},
191-197 (1992)}
\References{Se3}{A. Sengupta, {\it Quantum Gauge
Theory on Compact Surfaces}, Ann. Phys. {\bf 221},
17-52 (1993)}
\References{Se4}{A. Sengupta, {\it Gauge
Theory on Compact Surfaces}, preprint (1993)}
\References{Se5}{A. Sengupta, {\it The

Semiclassical Limit for $SU(2)$ and $SO(3)$
Gauge Theory on the Torus}, preprint (1994)}
\References{W1}{E. Witten, {\it On Quantum

Gauge Theories in Two Dimensions}, Commun. Math.
Phys. {\bf 141}, 153-209 (1991)}
\References{W2}{E. Witten, {\it Two Dimensional
Quantum Gauge Theory revisited}, J. Geom.
Phys. {\bf 9}, 303-368 (1992)}

\bye